\documentclass[12pt]{article}

\usepackage{epsf}

\newcommand{\eeq}{\end{equation}}

\newcommand{\beq}{\begin{equation}}
\newcommand{\beql}{\begin{eqnarray}}
\newcommand{\eeql}{\end{eqnarray}}

\begin{document}

\title{Volatility of an Indian stock market : A random matrix approach
}

\author{ Varsha Kulkarni$^1$  and Nivedita Deo$^{1,2}$\\
\\
$^1$Department of Physics and Astrophysics\\University of Delhi, Delhi 110007, India.\\
\\
$^2$Abdus Salam International Center for Theoretical Physics\\Trieste, Italy
}
\date {}
\maketitle

We examine volatility of an Indian stock market in terms of
aspects like participation, synchronization of stocks and
quantification of volatility using the random matrix approach.
Volatility pattern of the market is found using the BSE index for
the three-year period 2000-2002. Random matrix analysis is carried
out using daily returns of 70 stocks for several time windows of
85 days in 2001 to (i) do a brief comparative analysis with
statistics of eigenvalues and eigenvectors of the matrix $C$ of
correlations between price fluctuations, in time regimes of
different volatilities. While a bulk of eigenvalues falls within
RMT bounds in all the time periods, we see that the largest
(deviating) eigenvalue correlates well with the volatility of the
index, the corresponding eigenvector clearly shows a shift in the
distribution of its components from volatile to less volatile
periods and verifies the qualitative association between
participation and volatility (ii) observe that the Inverse
participation ratio for the 'last' eigenvector is sensitive to
market fluctuations (the two quantities are observed to anti
correlate significantly) (iii) set up a variability index, $V$
whose temporal evolution is found to be significantly correlated
with the volatility of the overall market index.

\section{INTRODUCTION}

Various physical phenomena occurring in space and time (like
Brownian motion, turbulence, chaos) have recently found
application in the study of dynamics of financial markets.
Financial time series originate from complex dynamical processes,
sometimes accompanied by strong interactions. The nature of
underlying interactions in a stock market is not known much the
same as in complex quantum systems. The interesting fact embedded
in stock market dynamics is that the daily prices on a stock
market are not formed by forces of conventional demand and supply.
The market may be considered as a gigantic complex dynamical
system of millions of transactions such that the traders strike
equilibrium prices. Every trader is an own-profit-maximizing agent
and her decision or range is not dependent on previous
transactions but on the time evolution of current events. If the
events occur randomly, prices will be random. In other words
knowing the course of a stock's price is of no consequence for
predicting its future [1]. Synchronization of the dynamics of
pairs of stochastically fluctuating stock prices can be modeled
using a correlation matrix. Such matrices are studied in the
context of nuclear physics [2]. Random matrix models have been
widely used in explaining the overall behavior associated with
spectra and eigenfunctions of complex quantum systems such as
interacting many body systems- one of them being the Stock Market
[3-4]. Time series, such as the stock market indices are closely
linked to the evolution of a large number of interacting systems
or a complex evolving system, now increasingly being studied by
physicists. Price changes in any market are sensitive to the
information arriving in the market. Seasonal and political cycles
of events and rare events like catastrophes increase speculation
and uncertainty in the market leading to high fluctuations in
prices or a volatile situation. Volatility is basically a measure
of the market fluctuations. The question of interest then is
whether and how volatility affects the response of market
dynamics. It has been found [5] that linkages between stock market
indices have tightened during financial crisis or highly volatile
periods. That is to say there is an overall rise in stock index
correlations in highly volatile periods or the rise in
correlations is accentuated during bouts of volatility. It is also
believed that active participation of traders, that is, higher
frequency of dealings/transactions, increases speculation and
uncertainty and hence the market fluctuations. A highly volatile
situation is also associated with heavy exchange of information in
the market. A number of researchers [3,4,7] have applied the
methods of RMT to financial data and found interesting clues about
the underlying interactions. This paper is an attempt, to exposit
some observations that may throw light on volatility. The purpose
of this paper is two-fold. First, it attempts to understand
quantitatively the closely related aspects of volatility such as
synchronization and participation of stocks in the market using
random matrix technique and second, to show that this technique
may be used to set up a quantity which possesses a strong
predictive power for the volatility of the market. We start with a
brief empirical analysis of the BSE index and show the volatility
pattern. The next section deals with the various exercises carried
out using random matrix approach. We conclude by discussing our
observations in the last section. We show a more detailed analysis
of the BSE index in the Appendix.

\section{EMPIRICAL ANALYSIS OF BSE INDEX}

We first consider statistical properties of the time evolution of BSE index. We label the time series of index as $Y(t)$. We calculate volatility and find the p.d.f of index changes.

\subsection{Data Analyzed}

This section uses the daily indices of the Bombay Stock Exchange (BSE) for a period of 3 years between 2000-2002. BSE is the largest market in India consisting of stocks from various sectors. Indices are basically an average of actively traded stocks, which are weighted according to their market value. Trading is done five days a week in this market, and we consider the opening values of indices to be continuous by removing the holidays. Each year corresponds to 250 days of elapsed time, approximately, the total number of data points in this set is 750.

\subsection{Volatility}

Long-range correlation has been found in the amplitude of price changes [6]. The presence of long- range dependencies in absolute value of price changes points to the existence of a ``subsidiary stochastic process'' commonly called as Volatility. As the term may imply, volatility is a measure of fluctuations that occur in the market. Volatility can be estimated by various methods such as- calculating the standard deviation of price changes, by Bayesian methods, by averaging the absolute values of price changes in an appropriate time window etc. Statistical properties of volatility prove to be of vital practical significance as volatility is a key parameter in risk management. It is directly linked to the information arriving in speculative markets. Rare events or unanticipated 'shocks', seasonal changes, economic and political cycles (elections, the announcement of the budget in a country) all tend to accentuate fluctuations. A highly volatile period is marked by increased uncertainty and a greater propensity of traders to speculate and interact. This situation is generally accompanied by high prices and is considered good for any market in terms of efficiency and activity.

\noindent Computing Volatility:

Volatility, as mentioned earlier, gives us a measure of the market fluctuations. Intuitively we can say that a stock whose prices fluctuate more is more ``risky'' or ``volatile''. We may formalize this as : Let $Y(t-\Delta t)$, $Y(t)$, $Y(t+\Delta t)$, $\ldots$ be a stochastic process where $Y(t)$ may represent prices, indices, exchange rates etc. The logarithmic returns $G(t)$ over time scale $\Delta t$ are


\begin{eqnarray}
 G(t) &=& \log(Y(t+\Delta t)) - \log(Y(t))\nonumber\\
      &\approx& {Y(t+\Delta t) \over Y(t)}
\end{eqnarray}
'$\Delta t$ ' refers to the time interval. In this case  $\Delta t = 1$ day.

Taking $Y$ to be the BSE index for the period 2000--2002, we plot its trend in figure~1. The figure shows a significant change in the value of index over the period of three years 2000--2002. The rate of change (decrease) appears to be more for the first 450 days than later. We may say that the Bombay stock exchange follows a long-term trend in the period considered in the sense that there is more uncertainty say four months in future than a month in future. The trend also reflects on the willingness to take risk on part of the traders; it seems the market was far more active in the year 2000 than 2001 or 2002. The downward trend maybe attributed mainly to the economic and/or political cycles in the country. There is a sharp dip in $Y$ near 9/11/2001 (425th day) after which the index rises and settles without much fluctuation. This is indicated in Figure~2.

\begin{figure}

\leavevmode \epsfxsize=4in \epsffile{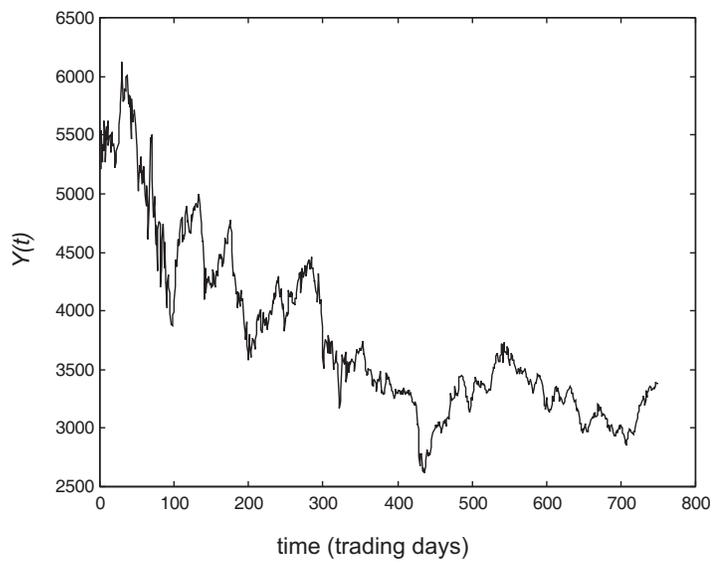}

\caption{BSE index plotted for all the days reported in the period 2000--2002. Total number of points plotted is 750.  A sharp dip can be seen around September 11th, 2001 (425th day in figure) when the index drops to the lowest.}

\end{figure}

\begin{figure}
\leavevmode \epsfxsize=6in \epsffile{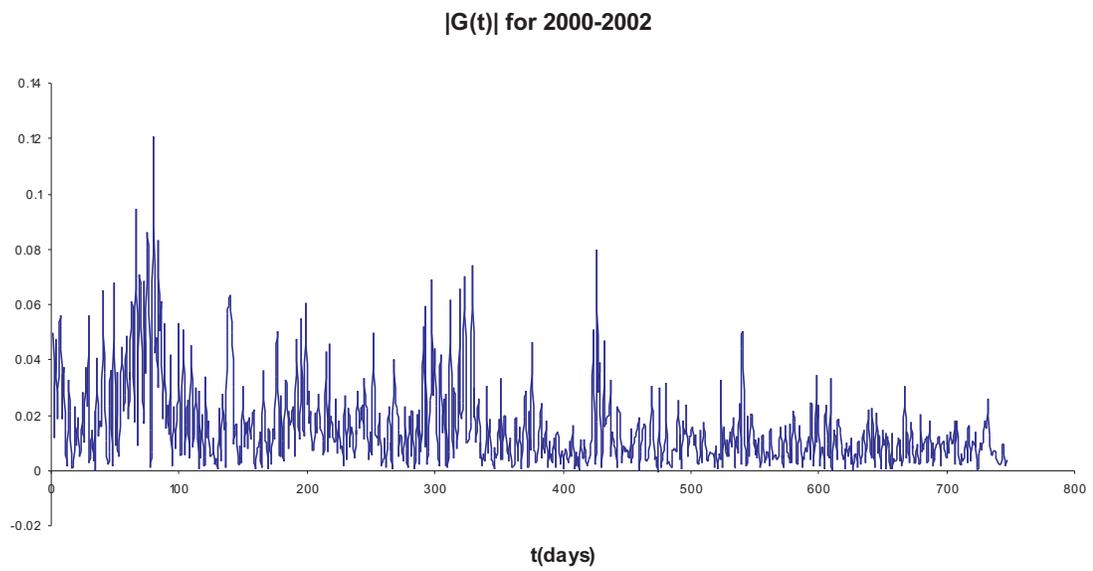}

\caption{Plot of absolute value of logarithmic returns of indices $|G|$ for the period 2000--2002.Large values of $|G|$ indicate crashes such as one on September 11th 2001 (425th day in the figure).}
\end{figure}

We quantify volatility, as the local average of the absolute value of daily returns of indices in an appropriate time window of $T$ days, as an estimate of volatility in that period

$$
  v(t) = {\displaystyle\sum_{t=1}^{T-1} |G(t)| \over T-1}
  \eqno(2)
  $$

We compute volatility for the three year period 2000--2002 by taking $T= 20$ days. The value of $T$ taken here corresponds to nearly a month as a month contains roughly 20 trading days in BSE. However, the results here may present some inaccuracy as the best estimation of volatility involves use of larger time periods. $\sum|G(t)|$ may be considered as a substitute for volatility or 'scaled volatility' in future.
\begin{figure}[h]

\leavevmode \epsfxsize=6in \epsffile{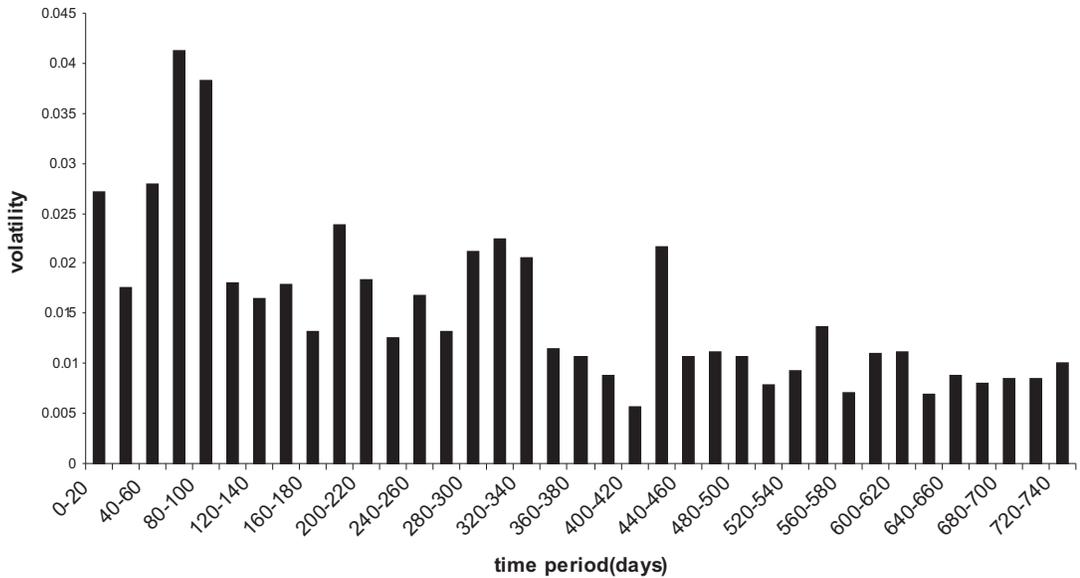}
\caption{Volatility,$v$ of BSE index for20-day periods between
2000-2002. Last ten days of the period have been ignored. The
period 420-440 days including the date of September 11th 2001
shows a sudden burst of activity.}
\end{figure}

Figure~3 shows the volatility of the market in the period
2000--2002. It is interesting to see from here three sub periods
(characterized by distinct volatilities respectively). Each year
corresponds to 250 days of elapsed time and we may divide the
period into three sub periods: I- 1-250 days (5.65), II-251-500
days (3.5), III- 501-750 days (2.25). We see that the year 2000
(1-250 days) was extremely active showing very high fluctuations
in the market and that regions of high volatility occur in
clusters showing consistently high fluctuations in say the first
200 days and more. Subsequently the fluctuations decrease in 2001
(251--500 days) . We find a more or less uneven fluctuating
pattern in this period. The period marked 420--440 shows a drastic
jump indicating that the event of 9/11 which happens to be
represented as the 425th day in the set, did increase the
volatility for that period. Obviously, this highly volatile period
does not show any precursors, as it was an unanticipated event
which rattled the market in that period, hence the jump reflects a
sudden impact on an otherwise quieter state of affairs. The event
of 9/11 was not long lasting as the last time period shows very
little fluctuation indicating a quiescent state in 2002 (501 days
onwards). These three time regimes characterized by distinct
patterns of volatility are of interest in order to observe the
volatility pattern of the market.

Since volatility is measured as the magnitude of the index changes, it may be worthwhile here to compare the return distributions for three time periods as considered before.

We calculate the probability distribution function (p.d.f), $P(Z)$ of daily returns, $Z$

$$
  Z_{\Delta t} = Y(t+\Delta t) - Y(t); \qquad \Delta t = 1 \mbox{day}
  \eqno(3)
  $$

We examine the nature of return distribution for three time periods of 250 days each as before. The distributions for all three periods confirm to a Levy stable regime (see Appendix). The three distributions (figure~4) differ in mean, standard deviation ($\sigma$) and the index of the distribution $\alpha$. While the periods: 251--500 and 501--750 do not differ much in maximum probability, the figure clearly shows the value of probability for small index changes, $P(Z-\delta z<Z<Z+ \delta z)$ for the period I is significantly lesser than (less than half) the corresponding values in the other two periods. However, period I shows a fatter tail, that is, it shows a higher probability for larger index changes than II, III. A more detailed analysis is shown in the appendix.

\begin{figure}
\leavevmode \epsfxsize=4in \epsffile{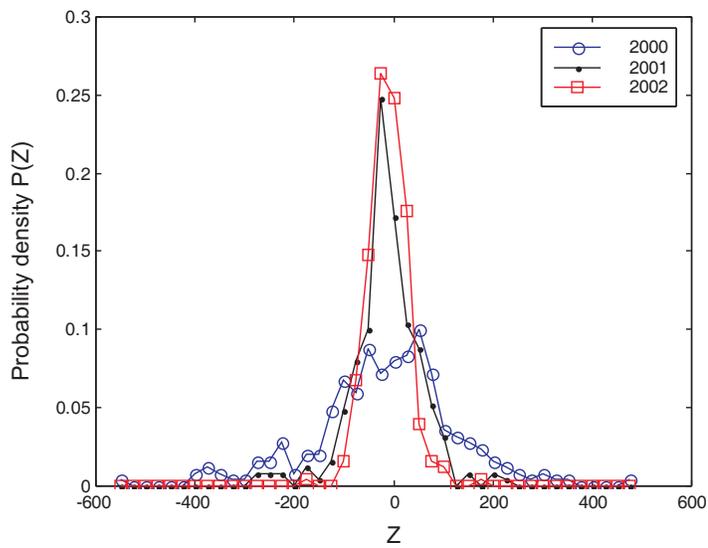}

\caption{The probability distributions of  daily index changes $Z$
are shown for three periods I, II, III corresponding to the years
2000, 2001, 2002 having mean$=-4.88$, $\sigma=141.58$,
$\alpha=1.51\pm 0.02$ for period I; mean=-2.76, $\sigma=69.96$,
$\alpha=1.38 \pm 0.01$ for period II; mean=0.53, $\sigma=39.01$,
$\alpha=1.57 \pm 0.02$ for period III. The value of $\delta z$ is
taken as 25.}
\end{figure}
\clearpage

\section {RANDOM MATRIX APPROACH}

Random Matrix Theory was developed by Wigner, Dyson and Mehta [2] in order to study the interactions in complex quantum systems. It was used to explain the statistics of energy levels therein. RMT has been useful in the analysis of universal and non universal properties of cross-correlations between different stocks. Recently various studies [3-4] have quantified correlations between different stocks by applying concepts and methods of RMT, and have shown that deviations of properties of correlation matrix of price fluctuations of stocks, from a random correlation matrix yield true information about the existing correlations.
While the deviations have been observed and studied in detail in the context of financial markets in earlier studies, we make a comparative analysis here, in the context of volatile versus less volatile situations from the point of view of correlations, participation of stocks in the market and try to quantify volatility in terms of the deviations.

\subsection {Data Analyzed and Constraints Involved}

As mentioned earlier BSE consists of stocks from various sectors and the market is operative five days a week. We must also mention that many of the stocks are not actively traded and hence not reported regularly in any period of time. Consequently they do not contribute much to the variations in stock price indices. Hence we consider here seventy stocks from largest sectors like chemical industry, metal and non-metal (diversified including steel, aluminum, cement etc). Since these stocks are actively traded throughout the year, we believe they would suffice to bring out our analysis of correlations in this section. Periods of analysis are confined to 280--500 days. The correlation matrices for this study have been constructed exactly along the same lines as in the earlier studies [3-4].

\subsection {Cross correlations}

We quantify correlations for $T$ observations of inter day price changes (returns) of every stock $i= 1, 2, \ldots, N$ as

$$
  G_i(t) = \log P_i (t+1) - \log P_i(t)
  \eqno(4)
  $$
where $P_i (t)$ denotes the price of stock $i$ and $t= 1, 2, \ldots, T-1$. Since different stocks vary on different scales, we normalize the returns as

$$
  M_i(t) = {G_i(t) - <G_i>\over \sigma}
  \eqno(5)
  $$
where $\sigma = \sqrt{<G_i^2> - <G_i>^2}$ is the standard deviation of $G_i$. Then the cross correlation matrix $C$, measuring the correlations of $N$ stocks is constructed with elements

$$
  C_{ij} = < M_i(t) M_j(t)>
  \eqno(6)
  $$
The elements of $C$ are  $-1 \le C_{ij} \le 1$. \\
$C_{ij} = 1$ corresponds to complete correlation\\
$C_{ij} = 0$ corresponds to no correlation\\
$C_{ij} = -1$ corresponds to complete anti correlation.

We construct the cross correlation matrix $C$ from daily returns
of $N=70$ stocks for two analysis periods of 85 days each (i)
280--365 days and (ii) 340--425 (see figure~3) marked with
distinct index volatilities respectively. The probability
densities of elements of $C$, $P(C_{ij})$ for both periods are
compared in figure~5. We see that the distribution for period (ii)
is more symmetric, implying a more or less equal extent of
positive and negative correlations than (i) which is characterized
by a more positive mean. The figure also suggests that there is a
de-concentration in higher levels of correlation magnitudes in a
less volatile period (ii) as compared to more volatile period (i).
A clear picture of existence of more pronounced correlations in
periods of high volatility is shown in Figure~7. The simple
correlation coefficient between the $<|C|>$ and volatility is
found to be 0.94 which is highly significant.
\begin{figure}[h]
\leavevmode \epsfxsize=4in \epsffile{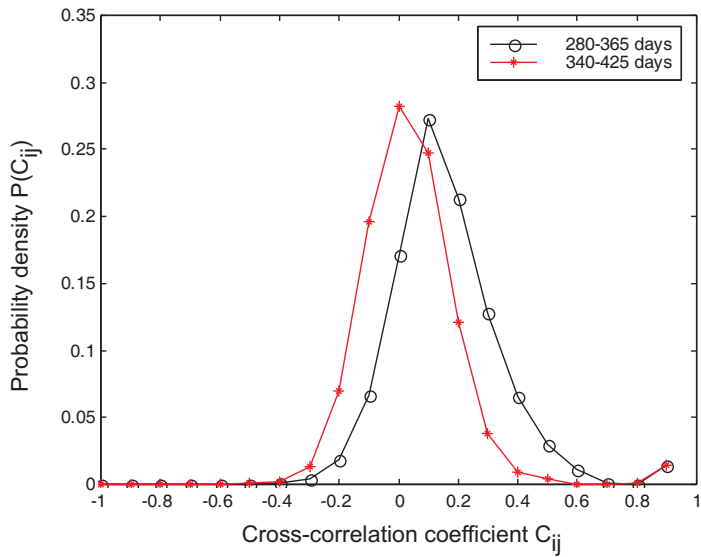}
\caption{Plot of the probability density of elements of
correlation matrix $C$ calculated using daily returns of 70 stocks
two 85 day analysis periods (i) 280-365 days and (ii) 340-425 days
with scaled volatilites of 1.6 and 0.82 respectively. We find a
large value of average magnitude of correlation $<|C|>= 0.22$ for
(i) compared to $<|C|>= 0.14$ for (ii).}
\end{figure}
\clearpage

\subsection{Statistics of Eigenvalues of C}

The eigenvalues of $C$ have special implications in identifying the true nature of the correlations. Earlier studies using RMT methods have analyzed $C$ and shown that $98\%$ of eigenvalues of $C$ lie within the RMT limits whereas 2\% of them lie outside [4]. It is understood that the largest eigenvalue deviating from RMT prediction reflects that some influence of the full market is common to all stocks, and that it alone yields ``genuine'' information hidden in $C$. The range of eigenvalues within the RMT bounds correspond to noise and do not yield any system specific information.

\noindent Eigenvalue Distribution of the Correlation Matrix

In order to extract information about the cross correlations from the matrix $C$, we compare the properties of $C$ with those of a random correlation matrix. $C$ is $N\times N$ matrix defined as

$$
   C = { G G^\top \over T}
   \eqno(7)
   $$

where  $G$ is an $N \times T$ matrix, $N$ stocks taken for $T$ days  and $G^\top$ denotes transpose of matrix $G$. We now consider a random correlation matrix

$$
  R = {A A^\top \over T}
  \eqno(8)
  $$
where  $A$ is $N\times T$ matrix with random entries (zero mean and unit variance) that are mutually uncorrelated. Statistics of random matrices such as $R$ are known. In the limit of both $N$ and $T$ tending to infinity, such that $Q= T/N(>1)$ is fixed, it has been shown that the probability density function Prm ($\lambda$) of eigenvalues of $R$ is given by

$$
  \mbox{Prm}(\lambda) = { Q \sqrt{(\lambda_+ - \lambda) (\lambda-\lambda_-)}
                                              \over 2\pi\lambda}
  \eqno(9)
  $$
for $\lambda$ lying in $\lambda_- < \lambda < \lambda_+$ where $\lambda_-$ and $\lambda_+$ are the minimum and maximum eigenvalues of $R$, respectively given by

$$
   \lambda_\pm = 1 + {1\over Q} \pm 2 \sqrt{{1\over Q}}
   \eqno(10)
   $$
We set up a correlation matrix $C$ from the daily returns of $N= 70$ stocks for $T=85$ days in the year 2001 for two periods (i) first excluding the data reported on the day- September 11, 2001- the 85th day being Aug 31st, and then (ii) sliding this window of 85 days to include the data reported on that day and beyond -the 85th day being September 18th.
Here $Q = 1.21$, and maximum and minimum eigenvalues predicted by RMT from (11) are 0.0086 and 3.6385.

\begin{figure}[h]
\leavevmode \epsfxsize=4in \epsffile{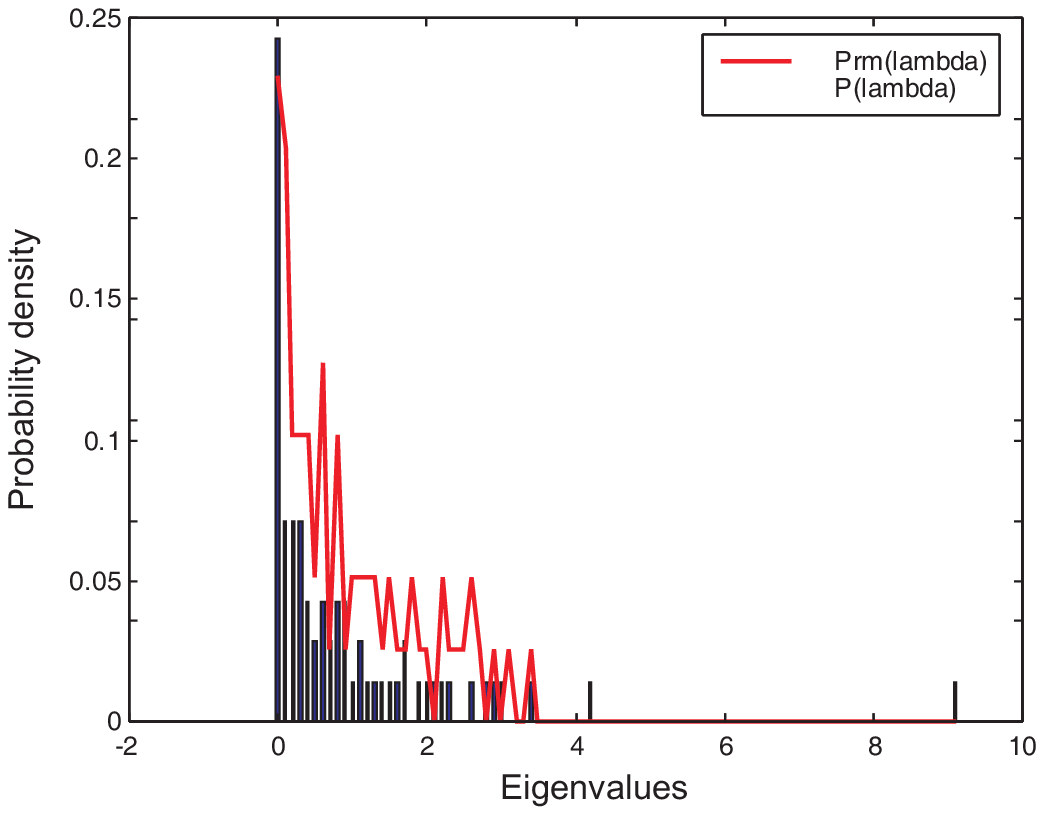}
\end{figure}

\begin{figure}[h]
\leavevmode \epsfxsize=4in \epsffile{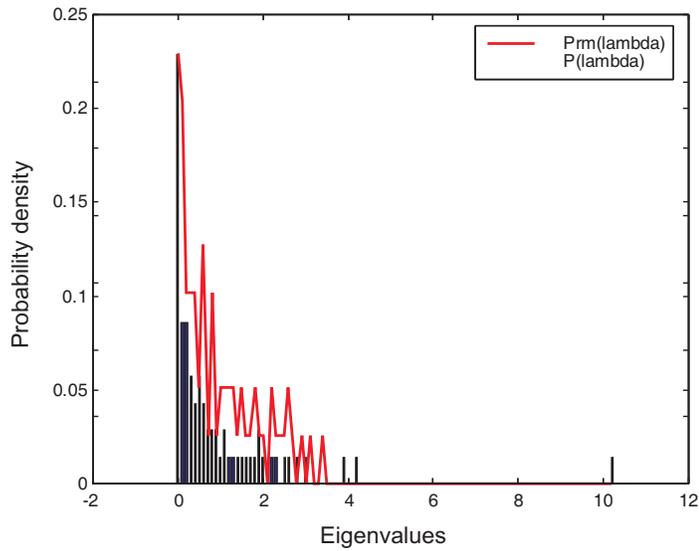}
\caption{Probability density of eigenvalues is shown by bars for a period considered (i) 334-419 before 9/11/2001 and having a volatility (scaled) of 0.8 (Top) and (ii) 346-431 including 9/11/2001 and having a volatility (scaled) of  0.9 (Bottom). A comparison is made with the probability density of eigenvalues of a random matrix R of the same size as C, shown by the solid line. The number of deviating eigenvalues is 4 in (i) and 6 in (ii). $\lambda_+$ for (i) is 9.17 and for (ii) is 10.28. }
\end{figure}
\clearpage

Figure~6 indicates and increased deviation in case (ii) as
compared to (i) in terms of both number of eigenvalues lying
outside RMT range and the magnitude of the maximum eigenvalue. In
the first case of a non-perturbed correlation matrix, 4
eigen-values lie outside RMT bounds; 2 larger than $\lambda_+$ and
2 smaller than $\lambda_-$. The largest eigenvalue is 9.17. In
case of a perturbed correlation matrix (ii), we find 6
eigen-values deviating from RMT limits; 3 larger than $\lambda_+$
and 3 smaller than $\lambda_-$. The maximum eigenvalue is 10.28.

\noindent Trend of Largest Eigen-values

Since the largest eigen-value is believed to represent true information about the correlations between stocks and is indicative of an influence common to all stocks, we wish to see the variation of the same as we move from a no-shock, quiescent period to the one hit by the shock of 9/11.
Here we start by setting up $C$ using daily returns of $N=70$ stocks for fixed but progressing time periods of length $T=85$ days. We look at the largest or ``deviating'' eigenvalue in the eigenvalue spectrum of $C$. The trace of the correlation matrix is preserved throughout, $\hbox{Tr} (C)=N$. The closer the maximum eigenvalue is to the trace, more information it contains and more correlated the prices would be. Variation of largest eigenvalue is seen by considering $N=70$ stocks for time windows of 85 days advanced each time by two days. Labeling the first and last day of all periods as $t_f$  and $t_l$ respectively, we set up $C$ as

$$
   C(t_f, t_l) = C(280 +j, 280+j+85)
   \eqno(11)
   $$
where $j= 0,2,4,6, \ldots, 134$.

Result of this exercise is shown in figure~7. We observe, a
decrease in the magnitude of largest for time periods spanning
280--425 days after which it is more or less constant. The largest
eigenvalue is found to be strongly correlated with volatility of
the BSE index (simple correlation coefficient is found to be 0.94)
for all times considered. We study the impact of 9/11 shock
(figure~8) by carrying out a similar exercise, taking $j= 0,1,2,3,
\ldots, 26$. The aftermath of the event can be seen in the sudden,
impulsive rise in the maximum eigenvalue around September
13th,18th, indicating that the impact was localized in time.

\begin{figure}[h]
\leavevmode \epsfxsize=4in \epsffile{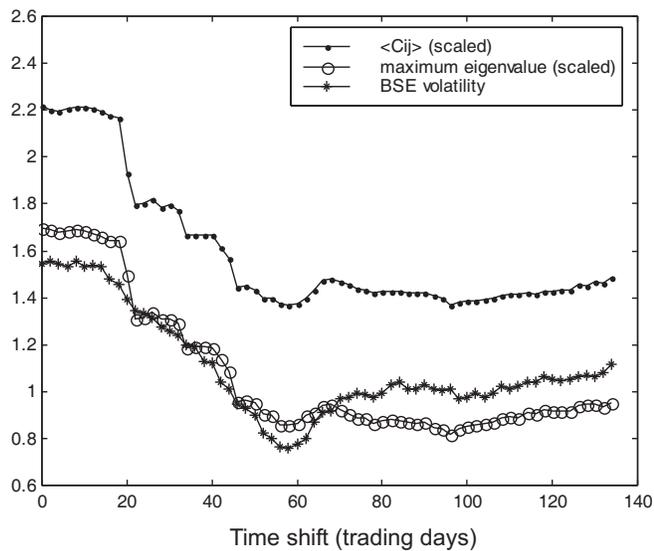}
\caption{Variation of largest eigenvalue and $<|C|>$, with the time shift, $j$. Analysis period is confined to period II. First $j$ was increased in steps of 2 days each time to span a total time of 280-500 days (see Figure~3). Volatility has been scaled for convenience. A minute exercise was carried out in Figure~8 by advancing the time windows in steps of 1 day each time, spanning a total time of 333-444 days in order to capture the impact of the 9/11 shock.  The horizontal axis shows the last day of all the time periods.}
\end{figure}

\begin{figure}[h]
\leavevmode \epsfxsize=6in \epsffile{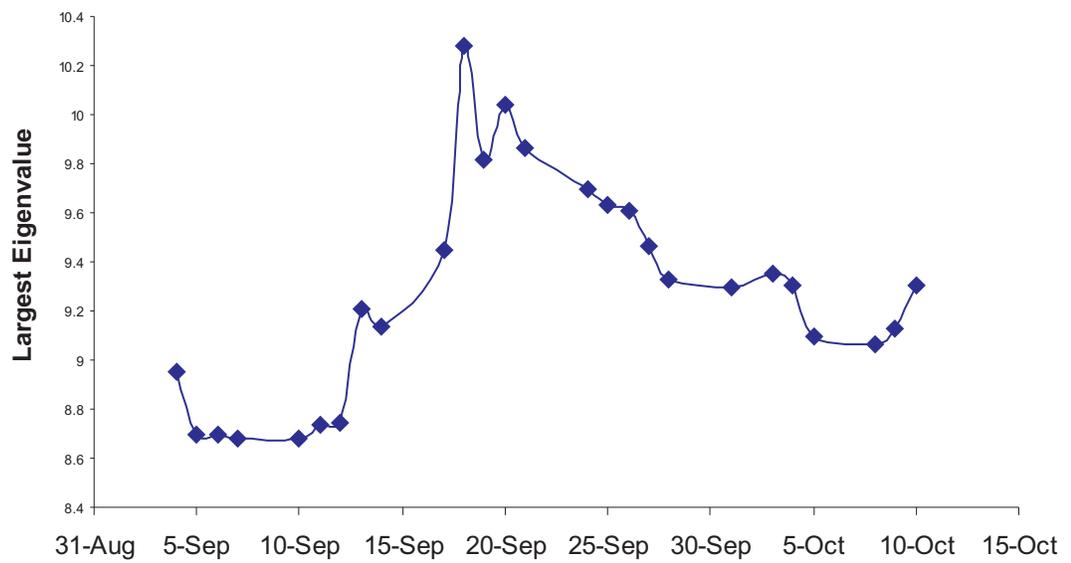}
\caption{Variation of largest eigenvalue with the time shift, $j$. Analysis period is confined to period II. First $j$ was increased in steps of 1 days each time to span a total time of 333-444 days in order to capture the impact of the 9/11 shock. (see Figure~3). Volatility has been scaled for convenience. The horizontal axis shows the last day of all the time periods.}
\end{figure}

\clearpage

\subsection {'Last' Eigenmode and the Variability Index}

The eigenstates of $C$ deviating from RMT predictions bring out the collective response of the market to perturbations. Collective motion of all the assets in the portfolio is significantly high, or the stocks are highly correlated in regimes marked by occasional or persisting bursts of activity. The degree of such synchronization is indicated by the eigenvector corresponding to the largest eigenvalue, through the evolution of its structure and components (in sub sections a \& b). Finally in c, we try to quantify volatility in terms of the largest eigenvector to yield a strong indicator of variability.

\subsubsection {Distribution of Eigenvector Components}

We wish to analyze the distribution of the components of the eigenvector corresponding to largest eigenvalue and compare the distributions for three time periods characterized by different volatilities (i) 280-365  (ii) 340-425 (iii) 380-465 as shown in figure~9.

Figure~9 shows the distributions of components of $U^{70}$ are shorter and broader in less volatile regimes (ii,iii) than in a more volatile one(i). Although the maximum participation is more in distributions (ii),(iii)  the number of significant participants in their sets (components differing significantly from zero) far lesser than (almost half) that in (i). This is dealt with in the next sub section. In addition we find that the components of $U^{70}$ for period (i), all have a positive sign, which confines the distribution to one side. This finding has been interpreted previously [4] to imply that a common component of the significant participants of $U^{70}$ affects all of them alike. We also find that for all periods that follow (iii) which are relatively quiescent and not shown here, contain both positive and negative elements. This goes to show an interesting link between the strength of the 'common influence' and volatility. We may say 'collective' or 'ensemble-like' behavior is more pertinent to volatile situations rather than non-volatile ones.

\begin{figure}[h]
\leavevmode \epsfxsize=4in \epsffile{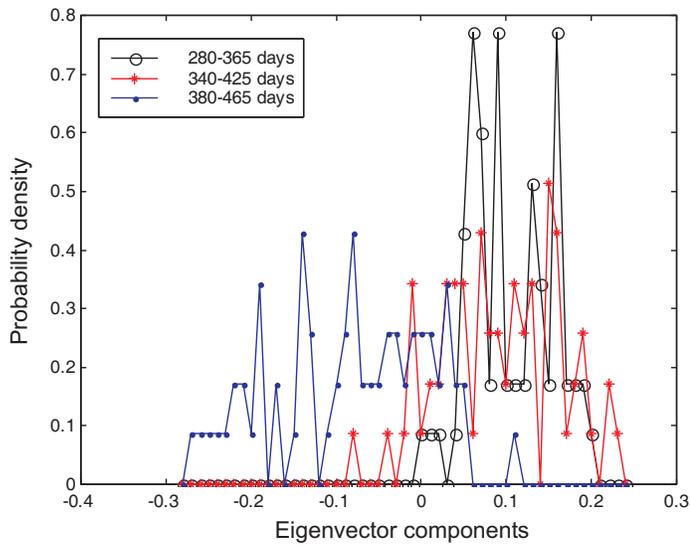}
\caption{Probability density of the eigenvector components for largest eigenvalue for three periods (i) 280-365 days (ii) 340-425 and (iii) 380-465 days marked by volatilities 1.6, 0.82, 0.99 respectively.  The plots are for $C$ constructed from daily returns of 70 stocks for $T=85$ days.}
\end{figure}
\clearpage

\subsubsection {Inverse Participation Ratio}

We analyze the evolution of the structure of the 'last' eigenstate, $U^{70}$ by evaluating the Inverse Participation ratio. The IPR quantifies the contribution of different components of eigenvector to the magnitude of an eigenvector. If $\nu_{i k}$,  $ i=1,2,\ldots, N$ be the components of eigenvector $U^k$ then IPR is given by

$$
  I_k = \sum_{i=1}^N \nu_{ik}^4
  \eqno(12)
  $$
Since IPR is actually the reciprocal of the number of eigenvector components that contribute significantly, if all components contribute identically, $\nu_{i k} = 1/\sqrt{N}$ then $I=1/N$. As before we set up a correlation matrix $C$ with $N=70$ stocks for $T=85$ days, each time shifting the time window forward in steps of 2 i.e. $j = 0,2,4,\ldots, 134$ spanning a period of 280--500 days as before The pattern of IPR (figure~10) indicates that the number of significant participants in $U^{70}$ decreases as we advance to less volatile periods. The IPR is closest to 0.0143 ( =1/70), the value we would expect when all components contribute equally, in the most volatile periods of the time span. The values of IPR deviate more and more from 0.0143 as we move to the less volatile periods. In fact the correlation between IPR and volatility was found to be equal to –0.63.

\begin{figure}[h]
\leavevmode \epsfxsize=4in \epsffile{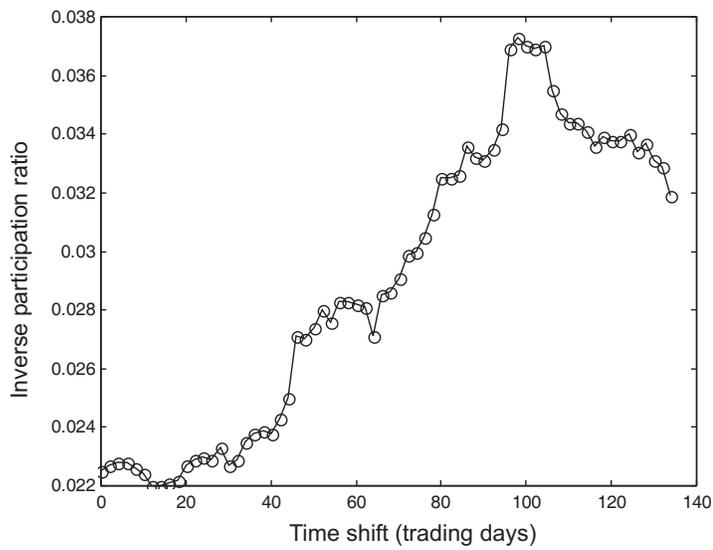}
\caption{Inverse participation ratio (IPR) for the eigenvector $U^{70}$ as a function of time. Results have been obtained from correlation matrix $C$ constructed from daily returns of 70 stocks for 68 time windows of 85 days each, progressed each time by 2 days spanning a time of 280-500 days. }
\end{figure}
\clearpage

\subsubsection {Variability index}

A yet another interesting feature brought out in the analysis of eigenvectors is the large-scale correlated movements associated with the last eigenvector, the one corresponding to largest eigenvalue. The average magnitude of correlations of prices of every stock m with all stocks $n= 1,2,\ldots,N$ is $<|C|>_m$ for $m= 1,2,\ldots, N$. is varied with the corresponding components of $U^{70}$ and $U^2$ (lying within the bulk) as shown in figures~11 and 12 respectively. While we find a strong linear positive relationship (shown in figure~11) between the two at all times for the $U^{70}$, the eigenvector belonging to the RMT range (figure~12) shows almost zero dependence. In this final sub section we make use of this dependence to set up a Variability index, which is strongly correlated with the variability of BSE index.
\begin{figure}[h]
\leavevmode \epsfxsize=4in \epsffile{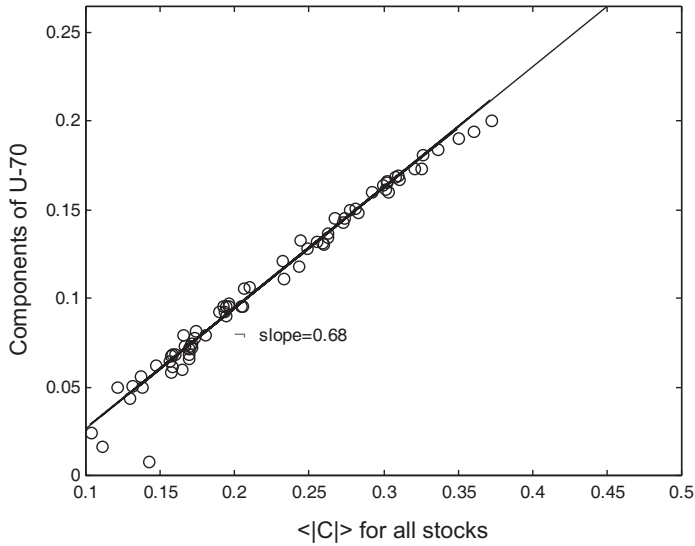}
\caption{Plot of the components of the eigenvector corresponding
to the largest eigenvalue with the extent to which every
individual stock is correlated in the market, denoted by
$<|C|>_m$. In this case, correlation matrix, $C$ was constructed
using daily returns of 70 stocks for the period 280-365 days. The
line obtained least square fitting has a slope$=0.68 \pm 0.01$.  }
\end{figure}
\clearpage
\begin{figure}[h]
\leavevmode \epsfxsize=4in \epsffile{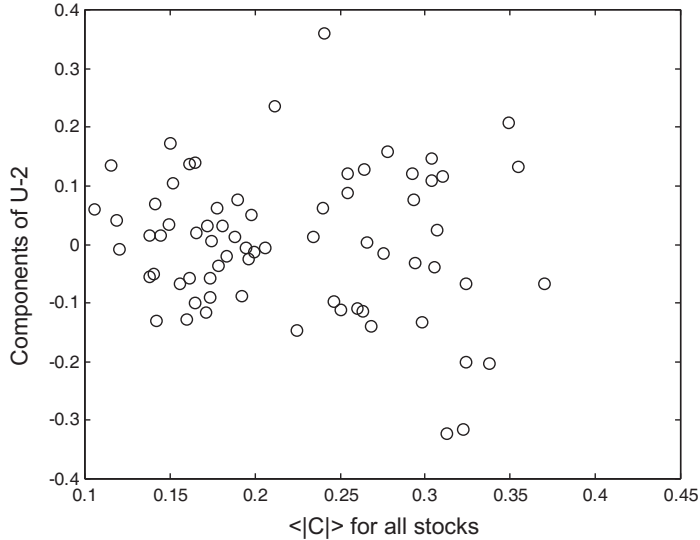}
\caption{Plot of the components of eigenvector $U^2$ associated
with an eigenvalue from the bulk of RMT, $\lambda_2$. The
variation shows no significant dependence between the two. The
picture is quite the same for successive time periods considered.}
\end{figure}

We define a projection vector $S$ with elements $<|C|>_m$ where $m=1,2,\ldots,70$, as calculated before. We obtain a quantity $X_m(t)$ by multiplying each element $S_m$ by the magnitude of the corresponding component of $U^{70}$ for each time window '$t$'.

$$
   X_m(t) = (U_m^{70})^2 S_m, \qquad m = 1, 2, \ldots, 70
   \eqno(13)
   $$
The idea is to weight the average correlation possessed by every stock m in the market according to the contribution of the corresponding component to the last eigenvector $U^{70}$, thereby neglecting the contribution of non-significant participants (close to zero) in $U^{70}$. The quantity $X$ in some sense represents the 'true' or 'effective' magnitude of correlations of stocks and the sum of such magnitudes are obtained as
$$
   V(t) = \sum_{m=1}^{70} X_m(t), \qquad \mbox{at time}\ t
   \eqno(14)
   $$
may be expected to reflect the variability of the market at that time. We call it the Variability index. We note from figure~13 that the variability index behaves remarkably similarly to the volatility of BSE index as the time window is slid forward. A highly statistically significant coefficient of correlation of 0.95 is obtained and a positive, linear relationship between the two can be seen in the plot of $V$ and BSE index volatility set out in figure~14. We thus find the relevance of the last eigenmode in quantifying the volatility of the overall market. Similar procedures have been carried out in other studies [8] in different contexts to verify the relevance of this last eigenvector.

\begin{figure}[h]
\leavevmode \epsfxsize=4in \epsffile{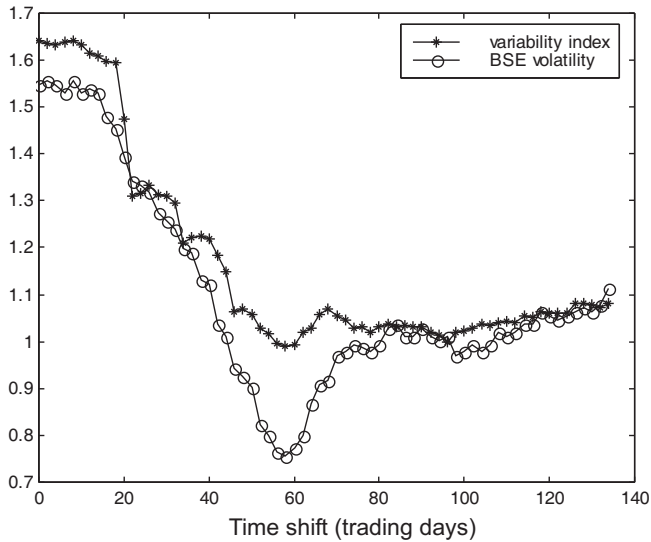}
\caption{Temporal evolution of the variability index, $V$ and the volatility of BSE index is shown upon suitable scaling. The results are obtained from correlation matrix $C$ constructed from daily returns of 70 stocks for 68 progressing time windows of 85 days each. The time was shifted in steps of 2 days each time and the time shift from the starting point is plotted on the horizontal axis.}
\end{figure}
\clearpage

\begin{figure}[h]
\leavevmode \epsfxsize=4in \epsffile{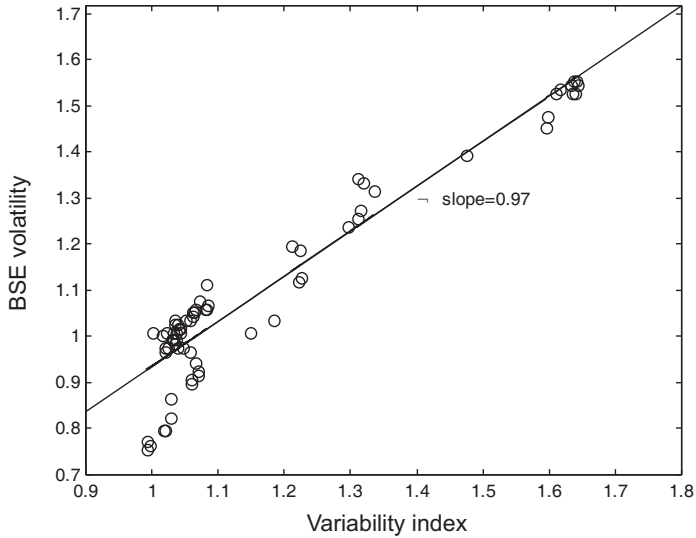}
\caption{The variability index, $V$ with the volatility of BSE
index approximates a linear fit with slope$= 0.97 \pm 0.04$}
\end{figure}

\section{CONCLUSION}

In this paper we study the volatility of the Bombay Stock Exchange
using the RMT approach. We find that the deviations from RMT
bounds are more pronounced in volatile time periods as compared to
the not so volatile ones in the context of Bombay Stock Exchange.
The largest eigenvalue, which is in some sense an index of 'true'
information in the entire market [7], is seen to be highly
sensitive to the trends of market activity. A comparison of
eigenvalue distributions for two analysis periods before and after
the event of 9/11, show that not only the number of eigenstates
deviating from RMT bounds but also the magnitude of the maximum
eigenvalue increases after the event. The simple correlation
coefficient between the two is
$\lambda_{\hbox{max}}$*BSEvolatility = 0.94. Analysis of the
correlation matrix $C$ as a function of time reveals a strong
dependence between the average of magnitude of elements of $C$ and
volatility, indicating highly synchronous movements of stocks in
highly fluctuating times or vice versa. A highly significant
correlation coefficient of 0.94 is observed here as well. The
eigenvector associated with the largest eigenvalue, the 'last'
eigenmode of $C$ has been enunciated in previous studies as a
collective response of the whole market to certain newsbreaks,
bursts of activity. We have tried to see its role in quantifying
the fluctuations. It has been understood previously by Plerou
et.al.[4] that if all the components of the eigenvector have the
same sign then there is some common component of the significant
participants that affects all of them with similar bias. The
probability density patterns of the components of $U^{70}$ show
that while the distribution in (i) is confined to the positive
values of participation, the other two have spread to the negative
side as well, indicating a gradual absence of existence of a
common influence on the components as we move from more volatile
period (i) to less volatile periods (ii,iii). Hence our finding
here may suggest that ensemble-like behavior is more prominent in
volatile situations than non-volatile ones. Further, the number of
significant participants in (ii, iii) falls to almost half that in
(i), a finding better exposited by the time evolution of the
Inverse participation ratio for components of $U^{70}$. A strong
anti-correlation between IPR and volatility (= -0.63) confirms the
existence of a positive association between the number significant
participants in $U^{70}$ with the volatility. It is verified that
the eigenvector $U^{70}$ indicates the extent to which the stock
movements are synchronized. We find a positive, linear
relationship between the extent to which all individual stocks
correlate or anti-correlate in the market $(<|C|>_m,
m=1,2,\ldots,N)$ and the corresponding elements of $U^{70}$.
Finally we investigate how this may lead to a quantification of
variability of the market by taking the product of $<|C|>_m$ with
squares of corresponding elements of $U^{70}$. The products for
all components may be put together as a sum to obtain a
Variability index, $V$. It is basically quantified as the sum of
correlations of individual stocks, each weighted according its
participation in $U^{70}$. Temporal evolution of $V$ and BSE index
volatility, have identical trends and there exists a highly
statistically significant correlation of 0.95 between the two. In
addition we find a close positive linear relationship between the
two. We may thus conclude that the 'last' eigenstate of the cross
correlation matrix can be set up usefully to obtain a quantity
that has statistically significant predictive power for the
variability of the market at any time.

\section* {ACKNOWLEDGMENTS}

We would like to thank Suresh Kulkarni, Raghav Gaiha and Vidyadhar
Mudkavi for the stimulating and fruitful discussions; Sachin
Maheshwari and Dheeraj Bharadwaj for helping us with the software.
S. Jain is thanked for encouragement and S. Mohapatra for
discussions.

\section* {APPENDIX}

\noindent{\bf Distribution of index changes}

\noindent Fluctuations in asset prices have been studied through Levy stable non-Gaussian model (Schulz~2001). Earlier studies [9] on S\&P500 index have shown the return distributions fit into the Levy stable regime with distribution index $\alpha$ : $0<\alpha<2$.

We calculate the p.d.f.  $P(Z)$ of daily returns, $Z$ defined by equation 3 in the text. We find the p.d.f. , $P(Z)$  is almost symmetric, sufficiently leptokurtic and possesses a narrow tail at the ends (figure~15). To characterize the functional form of the p.d.f., we see the variation of 'probability of return to origin' $P(Z-\delta z<Z<Z+\delta z)$ as a function of time interval $\Delta t$. This is done by taking various sets of returns $Z_{\Delta t} (t)$ corresponding to $\Delta t =1,2,3, \ldots,100$. We observe a power law decay (figure~15), which is consistent with L\'evy stable distribution.

In general, the 'probability of return to origin' is obtained as

$$
   P(Z=0) = { \Gamma(1/\alpha) \over \pi \alpha (\Delta t \gamma)^{1/\alpha}}
   \eqno(15)
   $$
The index of the distribution, $\alpha$ is the inverse of the power law exponent, and is found to be $1.66\pm0.01$. Probability of return to origin is 0.3623 and from (4) we get the scale factor of the distribution, $\gamma = 0.6662$.

\begin{figure}[h]
\leavevmode \epsfxsize=4in \epsffile{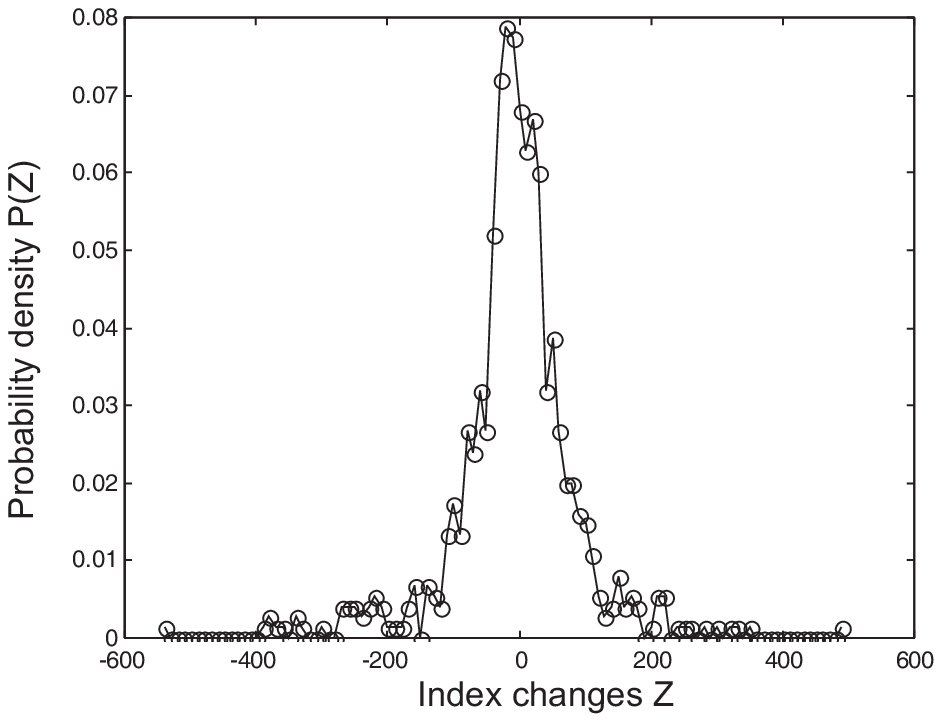}
\end{figure}
\begin{figure}[h]
\leavevmode \epsfxsize=4in \epsffile{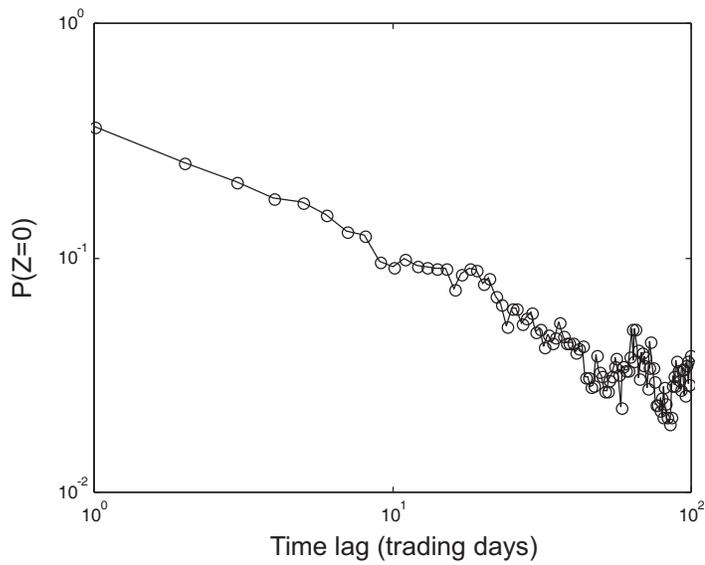}
\caption{Probability density function of index returns for the
period of 750 days between 2000 and 2002 (Top). Log-log plot of
'probability of return to origin', $P(Z-\delta z<Z<Z+\delta z$)
versus time lag, taking $\delta z=25$ shows a power law dependence
(Bottom). Straight line has slope$=-0.602\pm0.003$. The index,
$\alpha$ of the distribution in top figure is found to be
$1.66\pm0.01$. $P(-25<0<25)=0.3623$ for time lag of 1 day and
$\gamma=0.6662$.}
\end{figure}
\clearpage

Results of a more detailed analysis are shown in figure~16. The above analysis was carried out for 51 periods of 250 days each, starting from the first day of the set and advancing the time intervals each time by 10 days. We find the index $\alpha$ is almost constant $(1< \alpha < 2)$, but there is likeness between the variation of $\gamma$ and volatility of the BSE index. Probability of small index returns increases with every advance from highly volatile periods towards the relatively quiescent periods (see figure~3).
The plot of $\gamma$ and index volatility (figure~16) for the times considered shows that the data fit into two separate positive linear relationships. It is found that the rate of change in the periods of higher fluctuations, say till the first 340 days is 0.63, while it is 0.39 for then onwards. A possible explanation for this change of scale could be that probability of low index returns does not increase identically as the decrease in volatility. A highly significant correlation of 0.93 is obtained between $\gamma$ and volatility.

\begin{figure}[h]
\leavevmode \epsfxsize=4in \epsffile{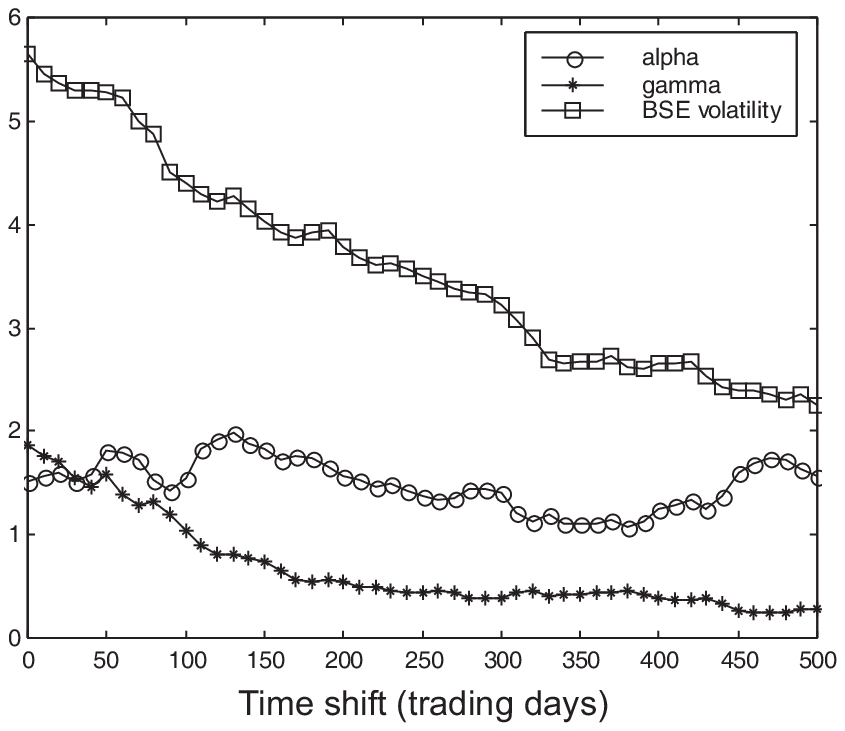}
\end{figure}
\begin{figure}[h]
\leavevmode \epsfxsize=4in \epsffile{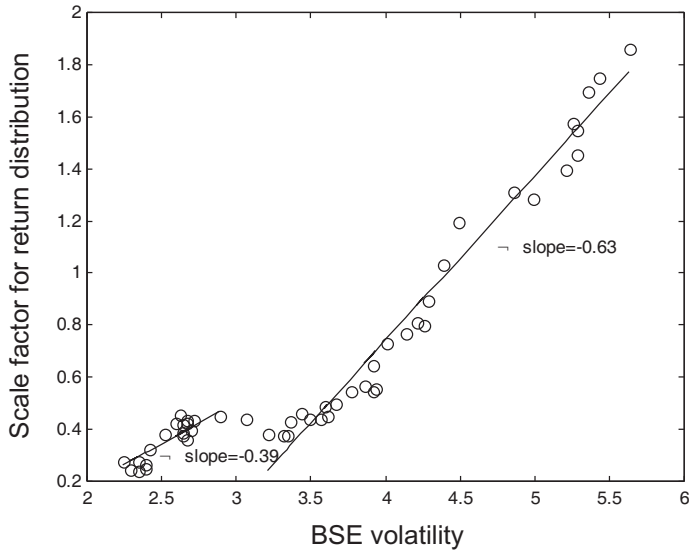}
\caption{Time dependence of volatility, distribution index $\alpha$ and scale factor $\gamma$ (Top). The observations were taken for 51 time periods of 250 days each. The time window was slid each time by 10 days to span the entire data. The horizontal axis shows the 1st day of all the 250 day periods considered with 45 lags each and $\delta z=25$. Bottom figure shows variation of
$\gamma$ and volatility. Two linear relationships with differing slopes are found. The less volatile periods (say 340-570 onwards) show a weaker dependence (slope=0.39) than the earlier more volatile periods (slope=0.63). In both the cases volatility has been scaled.
}
\end{figure}

\end{document}